\newcommand{\bm}[1]{\mbox{\boldmath$#1$}}
\documentstyle[multicol,prl,aps,epsbox]{revtex}

\input epsf

\title{Parisi States in a Heisenberg Spin-Glass Model in Three Dimensions }

\author{ F. Matsubara,  T. Shirakura$^1$, and Y. Baba }

\address{Department of Applied Physics, Tohoku University, Sendai 980-8579,
Japan\\
$^1$Faculty of Humanities and Social Sciences, Iwate University, 
Morioka 020-8550, Japan 
}

\date{ \today }

\begin{document}

\maketitle

\begin{abstract}

We have studied low-lying metastable states of the $\pm J$ Heisenberg model 
in two ($d=2$) and three ($d=3$) dimensions having developed a hybrid 
genetic algorithm. 
We have found a strong evidence of the occurrence of the Parisi states 
in $d=3$ but not in $d=2$. 
That is, in $L^d$ lattices, there exist metastable states with  
a finite excitation energy of $\Delta E \sim O(J)$ for 
$L \rightarrow \infty$, 
and energy barriers $\Delta W$ between the ground state and those 
metastable states are $\Delta W \sim O(JL^{\theta})$ with $\theta > 0$ 
in $d=3$ but with $\theta < 0$ in $d=2$. 
We have also found droplet-like excitations, suggesting 
a mixed scenario of the replica-symmetry-breaking picture and the droplet 
picture recently speculated in the Ising SG model.


\end{abstract}

\begin{multicols}{2}


Recently, the Heisenberg spin-glass (HSG) model in three dimensions ($d=3$) 
has been attacked a great interest, because evidence of the occurrence 
of SG phase transition at a finite, no-zero temperature ($T_{\rm C} \ne 0$) 
has been given in numerical studies in contrary to a common belief that 
any phase transition does not occur without some 
anisotropy\cite{BrayMY,Iyota2}. 
Kawamura and his coworkers took notice of chiralities of the spins 
and showed that a chiral glass (CG) phase transition occurs at 
$T_{\rm CG} \neq 0$, but the spin glass phase is still 
absent\cite{Kawamura1,Kawamura2,Kawamura3}. 
On the other hand, Matsubara et al. examined the stiffness 
at $T = 0$ and $T \neq 0$ 
of the $\pm J$ HSG model on the $L^3$ lattice with open boundaries 
and suggested that the SG phase transition will occur at 
$T_{\rm SG} \sim 0.19J$\cite{Matsu1,Matsu2}. 
They also obtained almost the same transition temperature by using 
different numerical methods, i.e., an aging effect\cite{Matsu3} and 
the divergence of the SG susceptibility\cite{Shira}. 
Recently, Nakamura and Endoh showed that, using non-equilibrium relaxation 
method, the CG phase transition and the SG phase transition occur at 
the same temperature of $T_{\rm SG} = T_{\rm CG} \sim 0.21J$\cite{Nakamura}. 
Quite recently, Lee and Young presented the same conclusion 
using a finite size analysis of the correlation length of the spins 
and chiralities\cite{Lee}.

An interesting question is, then, the nature of the SG phase of the HSG model. 
In the Ising SG (ISG) models in $d=3$, two pictures have been 
extensively discussed: 
the replica-symmetry-breaking (RSB) picture of Parisi\cite{Parisi} 
and the droplet picture of Fisher and Huse\cite{Droplet}. 
An important difference between the RSB and the droplet pictures concerns 
the nature of large scale excitations. 
In the RSB picture, there are many metastable states, which involve turning 
over a finite fraction of the spins and which cost only a finite 
energy even in the thermodynamic limit. 
The energy barriers between each of these metastable states and 
the ground state and also between those metastable states are infinite 
in the thermodynamic limit. 
Hereafter we call those metastable states {\it Parisi states}.  
By contrast, in the droplet picture, the lowest excitation which has linear 
spatial extent $l$ typically costs an energy of order 
$Jl^{\theta}$ with $\theta > 0$. 
Hence, in the thermodynamic limit, excitations which flip a finite 
fraction of the spins cost an infinite energy.


In this Letter, having developed {\it 
a hybrid genetic algorithm} (HGA) for systems with the XY and Heisenberg 
spins, we have studied the ground state and low-lying metastable states of 
the $\pm J$ HSG on finite lattices of $L^d$ ($d = 2$ and 3). 
{\it We have found the Parisi states in $d=3$ but not in $d=2$. } 
This finding is very important, because it gives a strong support of the 
presence of the SG phase in $d=3$. 
We have also found metastable states with droplet-like excitations 
near the Parisi states. 
We hope our findings will help to understand properties of the low 
temperature phase of the HSG model.


We start with the $\pm J$ Heisenberg model on the $d$ dimensional 
lattice of $L^d ( \equiv N)$ with periodic boundary conditions. 
The Hamiltonian is described by 
\begin{eqnarray} 
     H = - \sum_{\langle ij \rangle}J_{ij}\bm{S}_{i}\bm{S}_{j} 
\end{eqnarray} 
where $\bm{S}_{i}$ is the Heisenberg spin of $|\bm{S}_i| = 1$, 
and $\langle ij \rangle$ runs over all nearest-neighbor pairs. 
The exchange interaction $J_{ij}$ takes on either $+J$ or $-J$ with 
the same probability of 1/2. 

Before analyzing the model, we briefly mention properties of the 
Heisenberg SG model. First we should note that the model has 
the $SO(3) (\equiv SU(2)/\bm{C}_2)$ symmetry. 
So the ground state has a two-hold degeneracy. 
We call these two spin configurations as $G_1$ and $G_2$ and 
describe them $\{\bm{S}_i^{G_1}\}$ and $\{\bm{S}_i^{G_2}\}$. 
Note that, if $G_1$ is obtained, one can readily obtain $G_2$ by reversing 
all the spins of $G_1$, i.e., $\{\bm{S}_i^{G_2}\} = \{-\bm{S}_i^{G_1}\}$, 
and vice versa. 
The energy of the model is unchanged under any uniform rotation of the system. 
Then the spin configuration of the Heisenberg SG model is characterized 
by a set of pair-spin correlations $\{\bm{S}_i\bm{S}_j\}$. 
That is, the SG order parameter space is constructed by this set and 
the distance between two spin configurations $A$ with $\{\bm{S}_i^A\}$ and 
$B$ with $\{\bm{S}_i^B\}$ is described as 
\begin{eqnarray}
     S(A,B)    = \sqrt{\frac{1}{N^2}\sum_{i,j}(\bm{S}^{A}_i\bm{S}^{A}_j 
             - \bm{S}^{B}_i\bm{S}^{B}_j)^2}. 
\end{eqnarray}
Hereafter we take the ground state as the datum point and consider the 
distance $\Delta S$ of the spin configuration $A$ from the ground state 
spin configuration $G_1$ or $G_2$, i.e., 
$\Delta S \equiv S(A,G_1) = S(A,G_2)$. 
Note that, since the ground states $G_1$ and $G_2$ share the same point 
in the SG order parameter space, hereafter we omit the subscript. 
Note also that, if $\{\bm{S}_i^A\}$ is independent of $\{\bm{S}_i^G\}$, 
$\Delta S \sim \sqrt {\frac{1}{N^2} \sum_{ij}((\bm{S}^{A}_i\bm{S}^{A}_j)^2 + 
(\bm{S}^{G}_i\bm{S}^{G}_j)^2)} \rightarrow 
\sqrt{ \frac{2}{4\pi}\int \cos(\theta)^2 d\Omega} = \sqrt{2/3}$ 
for $L \rightarrow \infty$. 


Usually, the ground state of the HSG is searched by using a spin quench (SQ) 
method\cite{Kawamura1,Banaver}. 
However, the number $N_i$ of initial spin configurations which are 
needed to get the ground state increases rapidly as the size of the 
lattice increases. 
Here we develop a hybrid genetic algorithm (HGA) to search the ground 
state and low-lying metastable states. 
Starting with a population $N_p$ of random spin configurations (parents) 
$\{B_l\}$, new configurations (offsprings) are generated by recombination 
of different parents $B_l$ and $B_m\;(l \neq m)$ (quadruple crossover). 
Then some fraction $r$ of the spins are refreshed (mutation). 
This algorithm is hybridized with a local optimization 
(the SQ method) of the offsprings. 
The population $N_p$ is updated by replacing $3N_p/4$ parents with 
higher energy by $3N_p/4$ offsprings which are selected in order of 
the lower energy. 
This procedure is repeated for many times (generation number $N_g$). 
This algorithm is analogous to that used in the Ising SG model\cite{Pal}. 
In the Heisenberg SG, however, we should pay a special attention to 
optimize the interface energy between two parents  $B_l$ and $B_m$ 
when one generates the offsprings. 
This problem can be resolved by applying a uniform rotation to all the spins 
of one of the parents. Having used this algorithm, we have been able to obtain 
the ground state of the HSG model on larger lattices of $12\times12\times12$. 
The HGA also works well in the XY SG models.
Details of the HGA will be reported elsewhere.


Once the ground state $G$ with the energy $E_G$ 
are determined, one can obtain metastable states $A$ with $E_A$. 
We use the distance $\Delta S$ and the excitation energy 
$\Delta E (\equiv E_A -E_G)$ to distinguish different metastable states. 
We may use two different methods for searching those states: 
(i) the usual SQ method and (ii) the HGA. 
Both methods have their own merit. 
In the former, we can get different metastable states. 
However, we have to start with many different initial spin configurations 
to get low-lying metastable states, {\it e.g.}, $N_i\sim10^6$ for $L = 10$ 
in $d = 3$. 
On the other hand, the latter is appropriate to search the lowest 
metastable state in a given range of $\Delta S$. 
We are enough to give a smaller number of parents, 
{\it e.g.}, $N_p \sim 100$ for $L = 10$ in $d = 3$. 
In Figs. 1 (a) and (b), we present results of distribution of the metastable 
states of typical bond realizations (samples) in $d=3$. 
In fact, by using the two methods we can get the same lowest metastable state 
in a given range of $\Delta S$ (here $\Delta S > 0.4$). 
Usual excitation energies $\Delta E$'s increase with $\Delta S$ and also 
with $L$. 
A remarkable point is that $\Delta E$'s of the low-lying 
metastable states are much smaller than those usual excitation energies. 
In particular, we often see metastable states with finite $\Delta S$ 
which have a very small $\Delta E$ (see Fig.1(b)). 
This result implies the presence of the Parisi states. 
We will discuss this problem considering {\it an energy barrier between 
the ground state and those low-lying metastable states on the 
basis of a domain wall picture}. 
Another remarkable point is that, near the ground state, there are many 
metastable states the energy of which increases as $\Delta S$ increases. 
This result suggests the occurrence of some localized excitation the 
excitation energy of which increases as the scale of the excitation is 
increased\cite{Comm0}.
That is, the excitation will be droplet-like.   
The same is true for the excitations in metastable states near 
the Parisi states. 
We will get back to this point to consider properties of the 
low temperature phase. 

\begin{figure*}[bhtp]
\hspace{0.5cm} \psbox[scale=0.40]{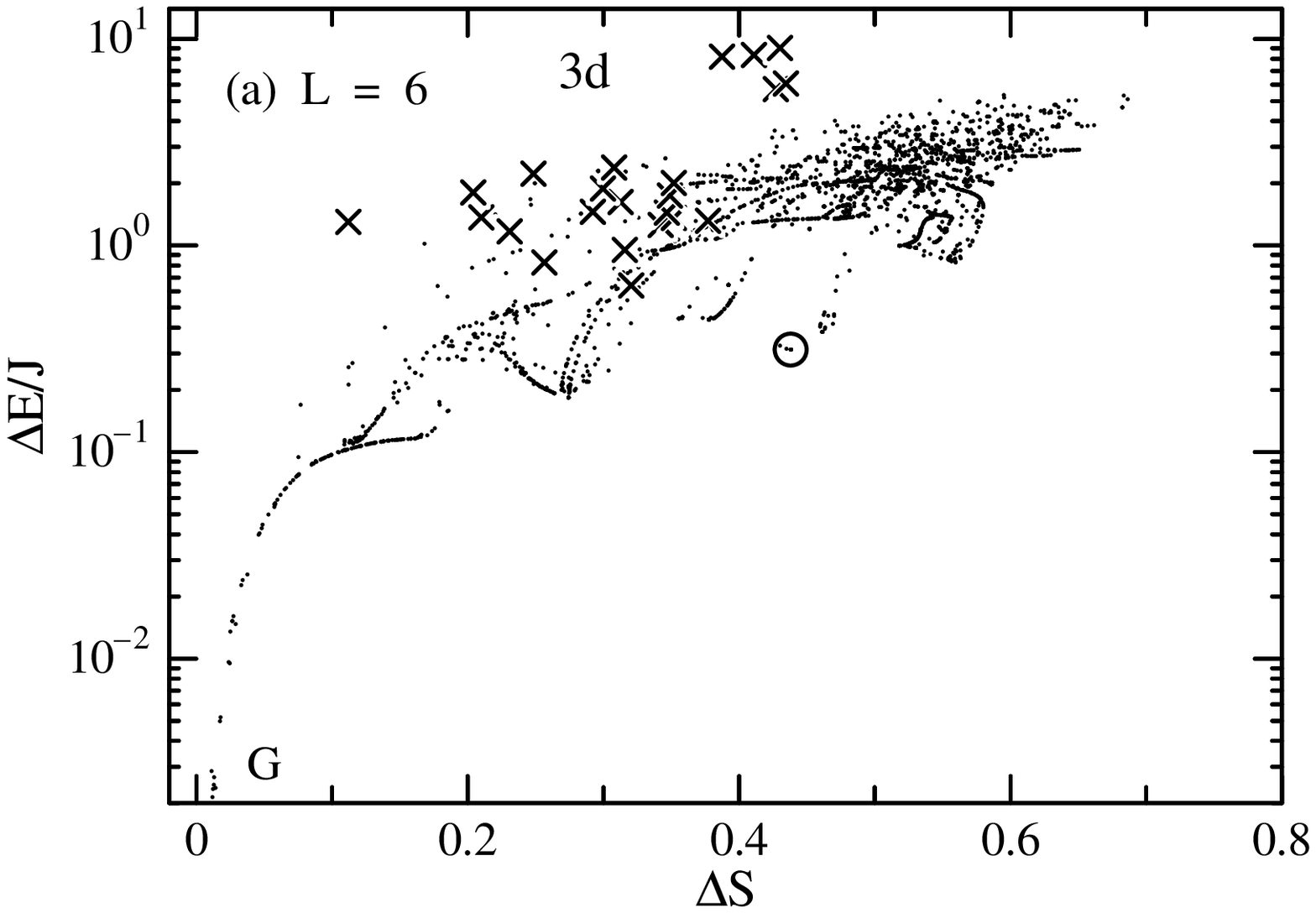}

\hspace{0.5cm} \psbox[scale=0.40]{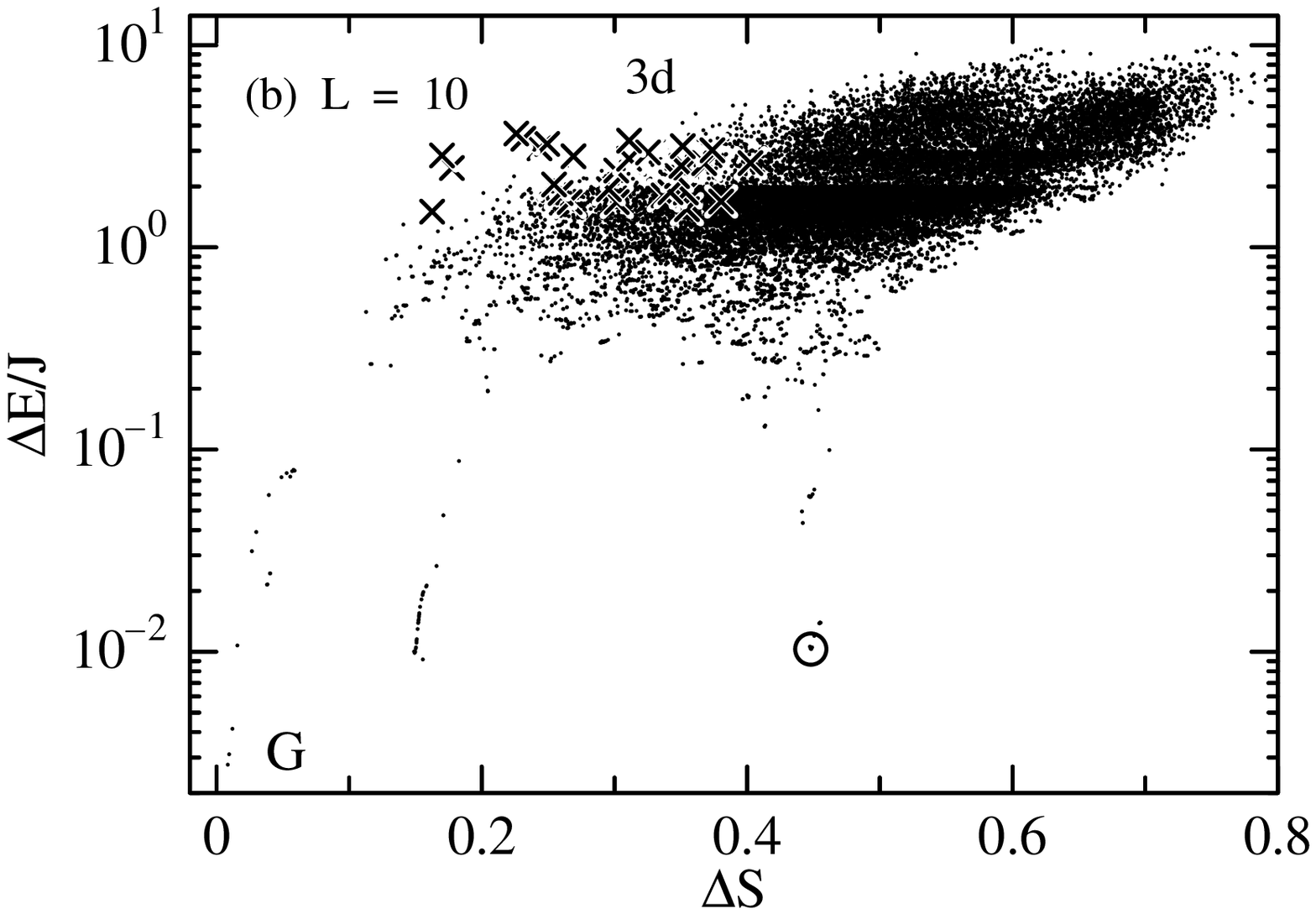}
\caption{
Distributions of metastable states (dots) in the ($\Delta S,\Delta E$) plane 
for typical samples with (a) $L = 6$ and (b) $L = 10$ in $d=3$ obtained 
by using the SQ method with $N_i \sim 2\times10^4$ ($L = 6$)
and $2\times10^6$ ($L = 10$). 
$\bigcirc$ indicates the lowest metastable state for $\Delta S > 0.4$ 
obtained by using the HGA and $\times$ denote the domain wall 
states described in the text. $G$ indicates the position of the 
ground state.
}
\label{fig:1}
\end{figure*}

Suppose that, in addition to the ground state $G$ with $E_G$, 
some low-lying metastable state $A$ is given. 
The domain wall energy $\Delta W$ between $G$ and $A$ may be 
estimated by the following procedure. 
(i) We divide the $L^d$ lattice into two parts ${H_1}$ and ${H_2}$, 
which are composed of $[(L+1)/2]$ and $[L/2]$ layers, respectively, 
where $[x]$ means the largest integer which does not exceed $x$. 
(ii) Fill ${H_1}$ with the spins of $G$ and ${H_2}$ with the spins of $A$, 
and apply a uniform rotation to all the spins on $H_2$ to minimize 
the interface energy. 
(iii) For each of $H_1$ and $H_2$, fix all the spins on the middle layer. 
Under this restriction, the SQ method is applied to get the spin 
configuration $W$ which gives the minimum energy $E_W$ of the $L^d$ lattice. 
(iv) We consider $W$ a domain wall state when $S(A,G)/4 < S(W,G) < S(A,G)$ 
and $S(A,G)/4 < S(W,A) < S(A,G)$ are satisfied, and define the domain wall 
energy $\Delta W = E_W - E_G$. 
The number of possible divisions of the $L^d$ lattice into ${H_1}$ and 
${H_2}$ is $d\times L$ and the chirality freedom of 2 exists. 
So we repeat this procedure $2dL$ times. 
In Figs.1(a) and (b), we have added $\Delta W$'s of these domain wall states 
between the ground state and the lowest metastable state for 
$\Delta S > 0.4$ indicated by the symbol $\bigcirc$. 
It is interesting to see that $\Delta W$'s are much higher than $\Delta E$. 


Now we consider the domain wall energies between the ground state 
and the lowest metastable state in a given range of $\Delta S$ for 
different samples. Here we consider the case $0.4 < \Delta S < 0.6$, 
because we are interested in large scale excitations. 
We denote the excitation energy of the lowest metastable state 
in this range as $\Delta E_0$. 
Our attention is paid to the minimum domain wall energy $\Delta W_{\rm min} 
(\equiv \min(\;\Delta W$'s$\;)$) for each of those samples. 
The calculation has been performed in $d =2$ and $d = 3$ by using the HGA. 
The linear sizes of the lattice are $L = 10 \sim 24$ in $d = 2$ 
and $L = 6 \sim 11$ in $d = 3$, and the numbers of the samples are 
$N_s = 1024$ in both $d = 2$ and $d=3$ except for the largest lattices  
($N_s = 512$ for $L = 24$ ($d=2$), and $N_s = 256$ for $L = 11$ ($d = 3$) ). 
The following numbers of $N_p$ and $N_g$ are chosen with a common mutation 
ratio $r = 0.4$. 
In $d = 2$, $N_p = 16, 32, 64, 128, 256$ for $L = 10, 12, 16, 20, 24$, 
respectively, and 
$N_g= 5$ for $L \leq 16$ and $N_g= 16, 32$ for $L = 20, 24$, respectively. 
In $d = 3$, $N_p = 16, 32, 64, 128, 256, 512$ for $L = 6, 7, 8, 9, 10, 11$, 
respectively, 
and $N_g = 5$ for $L \leq 8$ and $N_g = 8, 16, 32$ for $L = 9, 10, 11$, 
respectively. 
Note that, metastable states are not always found in this range of 
$\Delta S$ by using the HGA with $r = 0.4$\cite{Comm1}, 
especally in smaller lattices. 
In Fig. 2, we plot $(\Delta E_0, \Delta W_{\rm min})$ for different samples 
in $d = 3$. 
A remarkable point is that, as $L$ is increased, the distribution of 
$\Delta W_{\rm min}$ considerably spreads to the higher energy side, 
while the distribution of $\Delta E_0$ slightly shifts to the 
lower energy side. 
Note that, in $d = 2$, the distributions in both $\Delta E_0$ and 
$\Delta W_{\rm min}$ shift to the lower energy side with increasing $L$.  

\begin{figure*}[hbt]
\hspace{0.5cm} \psbox[scale=0.40]{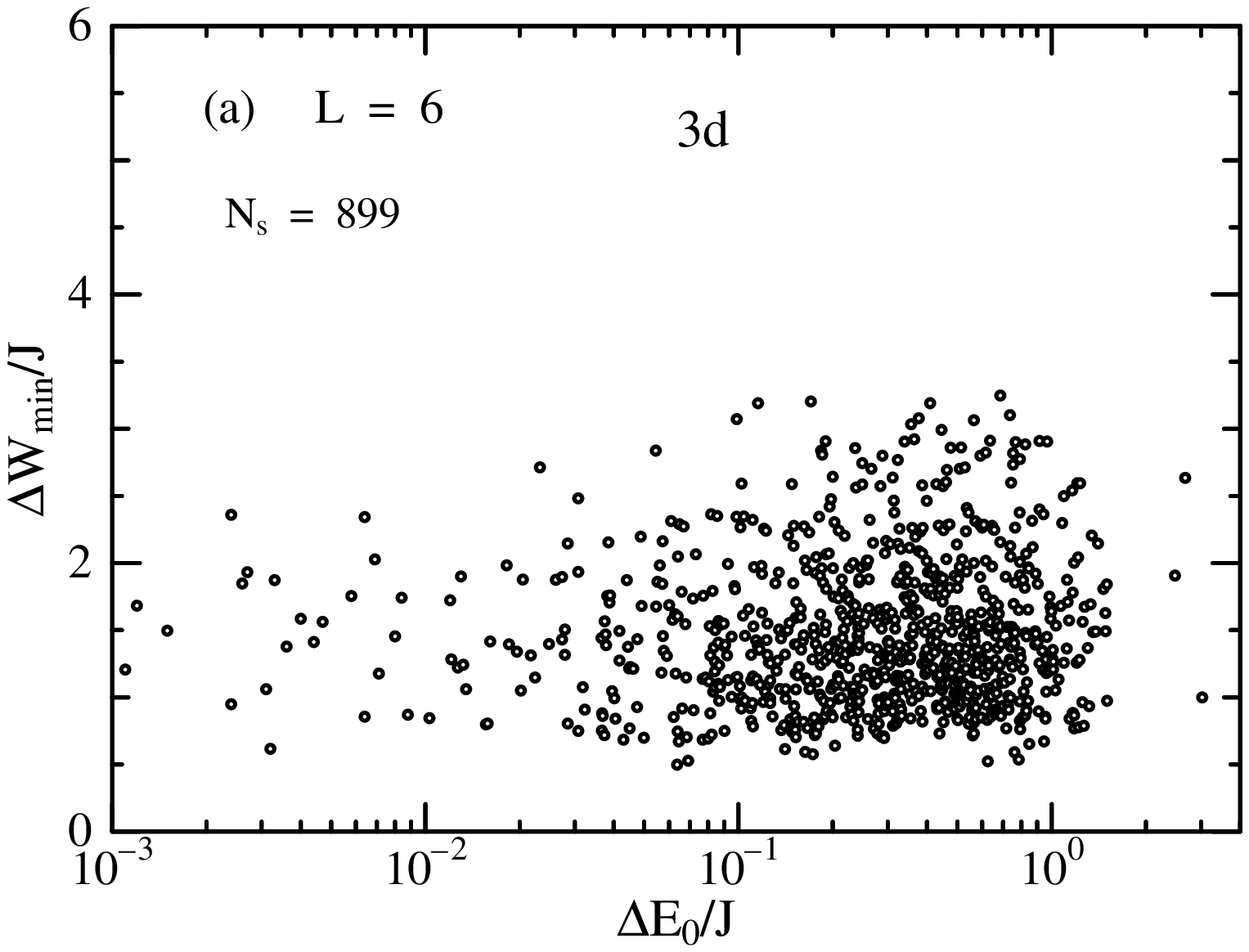}

\hspace{0.5cm} \psbox[scale=0.40]{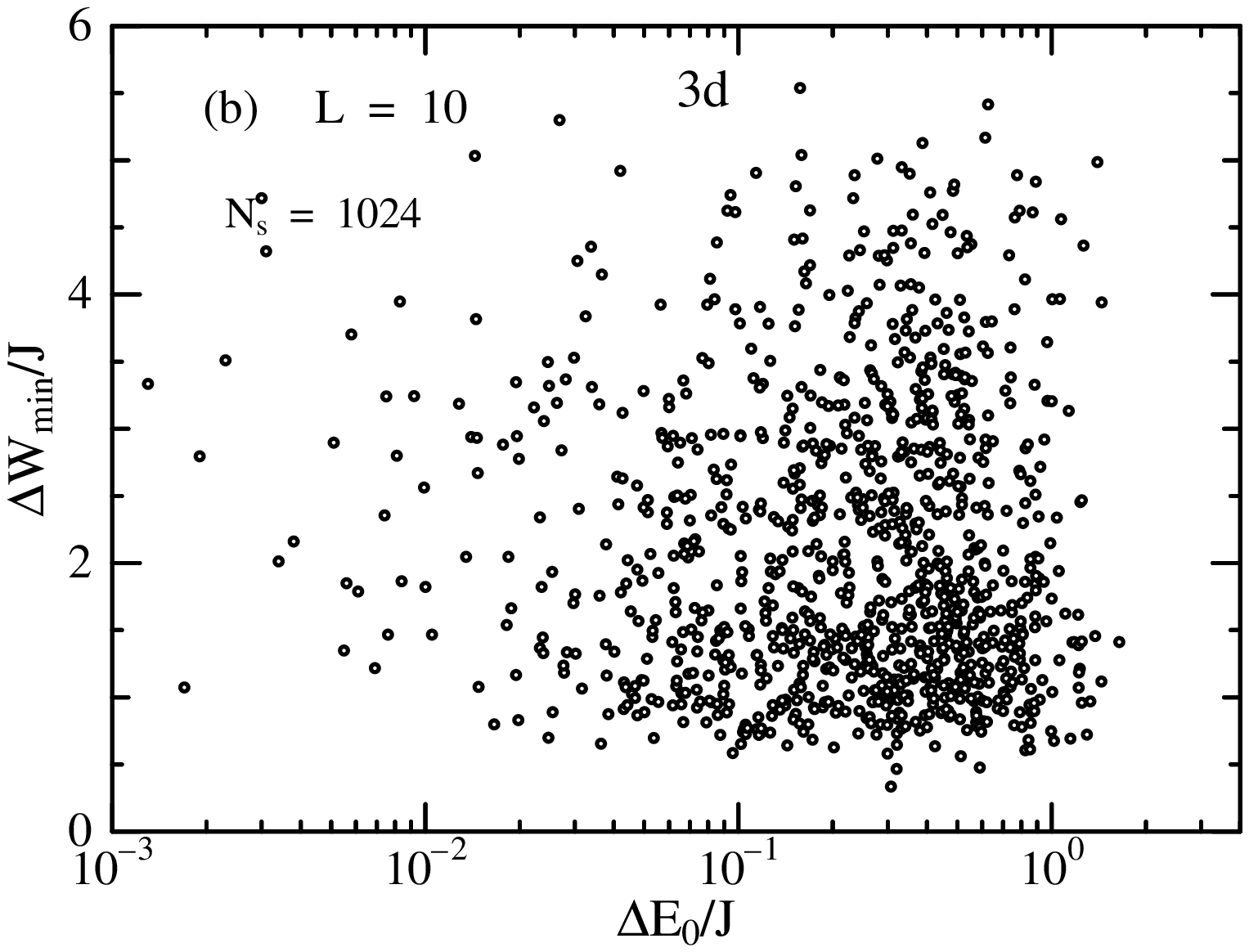}
\caption{
The lowest excitation energy $\Delta E_0$ vs the minimum domain wall energy 
$\Delta W_{\rm min}$ of different samples for (a) $L = 6$ and (b) $L = 10$ 
in $d = 3$. Here $N_s$ is the number of the samples in which some metastable 
states with $0.4 < \Delta S < 0.6$ are obtained among 1024 samples. 
}
\label{fig:2}
\end{figure*}

We calculate average values $\langle \Delta E_0 \rangle$ 
and $\langle \Delta W_{\rm min} \rangle$ over different samples and show them 
in Figs. 3 and 4 for $d=2$ and $d=3$, respectively, as functions of $L$. 
In $d=2$, in fact, $\langle \Delta E_0 \rangle$ and 
$\langle \Delta W_{\rm min} \rangle$ decrease with increasing $L$.  
These results clearly reveal the absence of the SG phase. 
By contrast, in $d=3$,  $\langle \Delta E_0 \rangle$ decreases 
slightly, and $\langle \Delta W_{\rm min} \rangle$ increases with $L$. 
We could fit data for larger $L$ as $\langle \Delta W_{\rm min} \rangle 
\propto JL^{\theta}$ with $\theta = 0.53 \pm 0.08$.
These results strongly suggest the presence of the Parisi states. 
That is, in the thermodynamic limit, there are metastable states 
which have a finite excitation energy $\Delta E_0$, 
probably $\Delta E_0 \ll J$, 
and which are separated from the ground state with an infinite energy barrier. 
The exponent $\theta$ is the measure of the domain wall height. 
It is interesting to find that the value of $\theta \sim 0.53$ 
is compatible with the stiffness exponent of $\theta = 0.4 \sim 0.8$ 
estimated recently\cite{Matsu1,Matsu2}.

%
\begin{figure*}[bht]
\hspace{0.5cm} \psbox[scale=0.40]{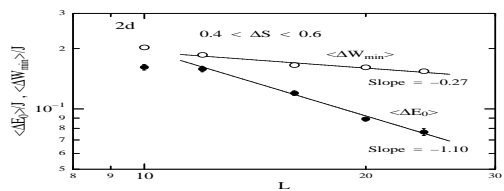}
\caption{
Average values of the lowest excitation energy $\langle \Delta E_0 \rangle$  
and the lowest domain wall energy $\langle \Delta W_{\rm min} \rangle$ 
in $d=2$ as functions of the linear lattice size $L$. 
}
\label{fig:4}
\end{figure*}
%

%
\begin{figure*}[htb]
\hspace{0.5cm} \psbox[scale=0.40]{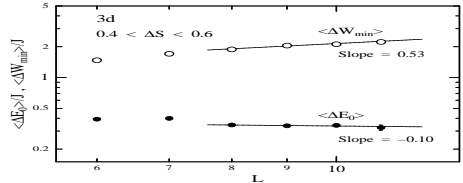}
\caption{
Average values of the lowest excitation energy $\langle \Delta E_0 \rangle$ 
and the lowest domain wall energy $\langle \Delta W_{\rm min} \rangle$ 
in $d=3$ as functions of the linear lattice size $L$. 
}
\label{fig:5}
\end{figure*}
%

In summary, we have studied low-lying metastable states of the $\pm J$ 
Heisenberg model in two ($d=2$) and three ($d=3$) dimensions 
having developed a hybrid genetic algorithm. 
We have found the Parisi states in $d=3$. 
Since the Parisi states occur in the SG phase, we suggest that 
the SG phase really occurs in the HSG in $d=3$. 
We have also found metastable states with droplet-like excitations 
near the Parisi states. 
So the SG phase of this HSG model will be characterized by a mixed 
scenario of the RSB picture and the droplet picture analogous to 
one\cite{Houdayer} that was speculated previously in the Ising SG 
model\cite{Mixed}. 
In this scenario, the system at $T = 0$ is located in one of the Parisi 
states and dynamical properties of the system at low temperatures may be 
described by using the droplet-like excitations thermally activated in 
that Parisi state. 
In the Ising model, aging dynamics of the SG model in $d=3$ has been 
reported to be well described by using the droplet picture\cite{Takayama}, 
although the model has large scale excitations with finite excitation 
energy\cite{Krzakala,Palassini}.  
We hope that the present findings stimulate studies of aging dynamics 
of the HSG as well as the equilibrium properties of the real SG systems.

\bigskip

The authors would like to thank Mr. S. Endoh and Dr. T. Nakamura 
for their valuable discussions. This work was financed by a Grant-in-Aid 
for Scientific Research from Ministry of Education, Science and Culture.

%

\end{multicols}


\begin{references}
%
\bibitem{BrayMY} A. J. Bray, M. A. Moore, and A. P. Young, 
          {\it Phys. Rev. Lett.} {\bf 56}, 2641 (1986).
%
\bibitem{Iyota2} F. Matsubara, T. Iyota, and S. Inawashiro, 
          {\it Phys. Rev. Lett.} {\bf 67}, 1458 (1991)
%
\bibitem{Kawamura1} H. Kawamura, {\it Phys. Rev. Lett.} {\bf 68}, 3785 (1992).
%
\bibitem{Kawamura2} H. Kawamura, {\it J. Phys. Soc. Jpn.} {\bf 64}, 26 (1995). 
%
\bibitem{Kawamura3} H. Kawamura, {\it Phys. Rev. Lett.} {\bf 80}, 5421 (1998).
%
\bibitem{Matsu1} F. Matsubara, S. Endoh, and T. Shirakura, 
        J. Phys. Soc. Jpn. {\bf 69}, 1927 (2000). 
%
\bibitem{Matsu2} S. Endoh, F. Matsubara, and T. Shirakura,
		J. Phys. Soc. Jpn. {\bf 70}, 1543 (2001).
%
\bibitem{Matsu3} F. Matsubara, T. Shirakura, and S. Endoh, 
        Phys. Rev. B {\bf 64}, 92412 (2001). 
%
\bibitem{Shira} T. Shirakura, F. Matsubara, and S. Endoh, 
        cond-mat/0009292; 
        F. Matsubara, T. Shirakura, S. Endoh, and S. Takahashi, 
        J. Phys. A, in press.
%
\bibitem{Nakamura} T. Nakamura and S. Endoh,
        J. Phys. Soc. Jpn. {\bf 71}, 2113 (2002). 
%
\bibitem{Lee} L. W. Lee and A. P. Young, Phys. Rev. Lett. {\it 90} 227203 
        (2003). 
%
\bibitem{Parisi} G. Parisi, Phys. Rev. Lett. {\bf 43}, 1754 (1979); 
        J. Phys. A {\bf 13}, 1101 (1980); {\it ibid.} {\bf 13}, 1887 (1980); 
        {\it ibid.} {\bf 13}, L115 (1980); 
        Phys. Rev. Lett. {\bf 50}, 1946 (1983).
%
\bibitem{Droplet} D. S. Fisher and D. A. Huse, J. Phys. A {\bf 20}, L997 
        (1987);
        D. A. Huse and D. S. Fisher, J. Phys. A {\bf 20}, L1005 (1987);
        D. S. Fisher and D. A. Huse, Phys. Rev. B {\bf 38}, 386 (1988).
%
\bibitem{Banaver} J. R. Banaver and M. Cieplak, Phys. Rev. Lett. {\bf 48} 
        832, (1982).
%
\bibitem{Pal} For example, K. F. P\'{a}l, Physica A {\bf 223} 283 (1996);
        and references therein. 
%
\bibitem{Comm0} We have also examined the spin structures of the metastable 
        states for $\Delta S \sim 0$ to confirm our speculation. 
        Result will be reported in a separate paper.
%
\bibitem{Comm1} Of course, we could find excited states in any range of 
        $\Delta S$ if we apply no mutation($r = 0$). 
%
\bibitem{Houdayer} 
        J. Houdayer and O. C. Martin, Europhys. Lett. {\bf 49}, 794 (2000).
%
\bibitem{Mixed} 
        This mixed scenario was proposed by Houdayer and Martin (ref.
         \cite{Houdayer}). 
        Immediately after that, the mixed scenario was redefined in somewhat 
        different manner and called as {\it TNT scenario} 
        (see refs. \cite{Krzakala} and \cite{Palassini}). 
        In the {\it TNT scenario}, a link overlap $q_l$ has been studied 
        together with the spin overlap $q_s$. We are currently calculating 
        $q_l$ to examine the {\it TNT scenario}. 
%
\bibitem{Takayama} T. Komori, H. Yoshino, and H. Takayama, J. Phys. Soc. Jpn. 
        {\bf 68}, 3387 (1999); {\it ibid.} {\bf 69}, 1192 (2000). 
%
\bibitem{Krzakala} 
        F. Krzakala and O. C. Martin, Phys. Rev. Lett. {\bf 85}, 3013 (2000).
%
\bibitem{Palassini} 
        M. Palassini and A. P. Young, Phys. Rev. Lett. {\bf 85}, 3017 (2000).
%
\end{references}
\end{document}